\documentclass[final,5p,times,preprint,twocolumn]{elsarticle}

\usepackage{amssymb}
\usepackage{amsmath}
\usepackage{subfigure}
\usepackage{lipsum}
\usepackage{flushend,cuted}
\usepackage{booktabs}  

\journal{Optics Communications}

\begin{document}

\begin{frontmatter}



\title{Bi-static LIDAR systems operating in the presence of oceanic turbulence}

\author[label1]{Olga Korotkova\corref{cor1}}
\author[label2]{Jin-Ren Yao}
\address[label1]{Department of Physics, University of Miami, Coral Gables, Florida 33146, USA}
\address[label2]{School of Physics, Harbin Institute of Technology, Harbin 150001, China}
\cortext[cor1]{Corresponding author: korotkova@physics.miami.edu}
\begin{abstract}
Optical turbulence occurring in the oceanic waters may be detrimental for light beams used in the short-link communication and sensing systems, and, in particular, in underwater LIDARs. We develop a theory capable of predicting the passage of light beams through the bi-static LIDAR systems, for a wide variety of optical waves, including partially coherent and partially polarized, and for a wide family of targets. Our theoretical framework is based on the Huygens-Fresnel integral adopted to random media and  optical systems described by the $4\times4$ ABCD matrices. The treatment of oceanic turbulence relies on the recently introduced power spectrum model of the fluctuating refractive-index [Opt. Express 27, 27807 (2019)] capable of accounting for different average temperatures of water. We first analyze the evolution of the second-order beam statistics such as the spectral density and the degree of coherence of the beam on its single pass propagation and then incorporate this knowledge into the analysis of the bi-static LIDAR returns. 
\end{abstract}




\end{frontmatter}


\section{Introduction}
\label{sec:intro}
\begin{figure*}
	\centering
	\includegraphics[width=1\textwidth]{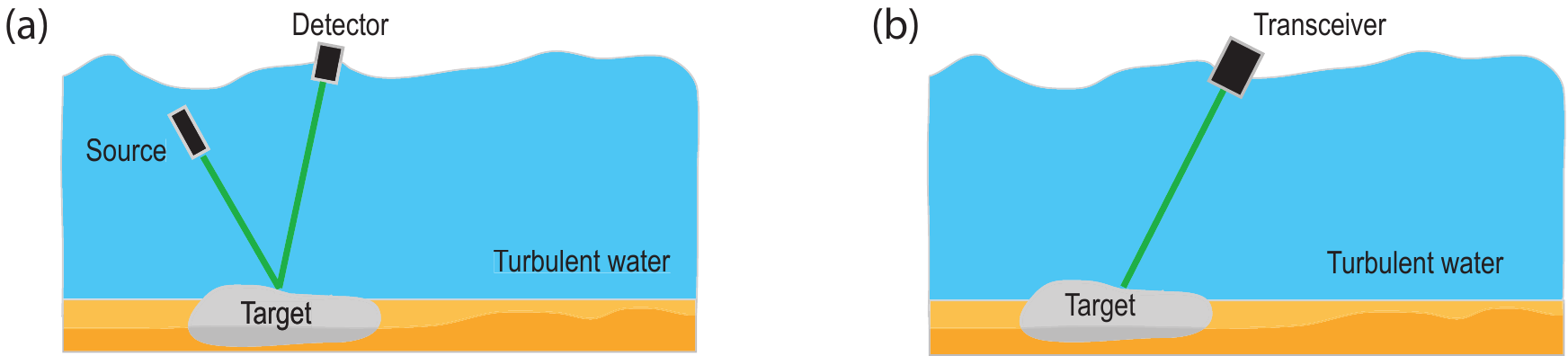}
	\caption{Bi-static and monostatic underwater LIDAR systems.}
	\label{fg:f1}
\end{figure*}

Light propagation through the double-pass links embedded in the oceanic water columns can be used in LIDAR technology either for the purpose of sensing the statistics of turbulence or for pointing and tracking of remote targets embedded into it \cite{AP}-\cite{Smith}. It also is the basis for the emerging retro-reflection modulation (RRM) communication technology \cite{RRM}. In this study we restrict ourselves to the analysis of target-sensing LIDARs operating in the clear-water channels, i.e., in the absense of the particulate content. One distinguishes bi-static/mono-static LIDAR links in which the transmitter and the receiver are spatially separated/co-located \cite{AP}. In the case of a bi-static link [see Fig. \ref{fg:f1}(a)] the optical wave can be analyzed sequentially: propagation through turbulence, interaction with target and propagation back to the receiving system, which substantially simplifies calculations and makes it possible to predict outcomes for a variety of sources, turbulent conditions and targets. For solving the direct propagation problem a $4\times4$ ABCD matrix method previously developed for atmospheric channels \cite{AtmLIDAR1}, \cite{AtmLIDAR2} can be readily adapted for oceanic channels. This method enables calculation of any second-order statistical properties of the returned beam, including its spectral density, coherence and polarization states, from the knowledge of these properties for the incident beam at the source plane. The required information about the target must include its geometry and the two-point, second-order height correlation properties of its surface. Without loss of generality, a Goodman-like target of circular shape, arbitrary size, curvature, having  Gaussian statistics and Gaussian correlation function of surface roughness can be modeled in. The inverse problem of finding the parameters of the target embedded in the turbulent medium, from comparison of the incident and the returned waves statistics, has also been tackled for the atmospheric propagation setting \cite{AtmLIDAR3}, \cite{AtmLIDAR4} and can also be readily adjusted for oceanic propagation problems. 

The situation is different for the mono-static LIDAR channels [see Fig.\ref{fg:f1}(b)] which can be formed for a variety of targets, including a mirror, a corner-cube retro-reflector, a Lambertian surface, etc. allowing at least a portion of the reflected beam intensity to propagate back to the transmitter along precisely the same path. Then, depending on the reflector's type, the return wave can exhibit either enhancement or suppression of the average intensity (and other statistics, e.g., the scintillation index) within a small area around the optical axis, the effect well known in atmospheric propagation as the Enhanced-BackScattering (EBS) \cite{Banach}, \cite{EBSreview}. The EBS portion of the beam is very stable which has been verified experimentally for a variety of air turbulence conditions \cite{Nelson}. \cite{JiaEBS}. This type of EBS must not be confused with backscattering from suspended particles: we assume here that it is entirely to do with turbulence. While the statistics of the return beam within the EBS area are very hard to predict, those outside of it are exactly the same as for the bi-static system of the same range. Hence, our results are also applicable for mono-static LIDARs outside of the EBS area.  In order to theoretically predict the EBS in the beam statistics, the correlations between the incident and return waves must be taken into account, in addition to all other calculations pertaining to the bi-static configuration. So far, the underwater EBS effect has been theoretically predicted only for the special cases of an incident plane and spherical waves \cite{KorEBSocean}, bounced by a corner-cube retro-reflector, but has not been experimentally confirmed, to our knowledge. We will also assume that while the source may be of electromagnetic nature, neither the ABCD system nor the target introduces polarimetric modulation, either deterministic or random, to the propagating light. If this occurs, the calculations must become much more involved, both in bi-static and mono-static scenarios \cite{OKsoresi}.

The statistics of optical waves interacting with the oceanic turbulence can be predicted from the knowledge of the power spectrum of the refractive-index fluctuations within the LIDAR channel. In a typical oceanic turbulence, two scalar fields, the temperature and the salinity are advected in the turbulent velocity vector-field resulting in two corresponding power spectra \cite{Hill1}, \cite{Hill2}, \cite{Thorpe}. These individual power spectra were first put together via a polynomial linearization procedure of Hill's model 1 by Nikishovs \cite{Nikishovs} (see also \cite{aY1} for a modified version). The oceanic power spectrum was recently reconsidered in Ref. \cite{ar01} where a very accurate numerical Hill's model 4 \cite{ar01} was analytically fitted within the wide range of the Prandtl/Schmidt numbers [3,3000] making it possible to consider the effects of average water temperatures in the range [0$^o$C, 30$^o$C], which covers practically all possible situations occurring within the oceanic portion of the boundary layer of Earth. Single-pass propagation of optical waves through the water turbulence with the Nikishovs' power spectrum has been explored in depth \cite{KorOceanRev}. In particular, the changes in the average intensity \cite{OP1}, spectrum \cite{OP2}, polarization \cite{OP3}, coherence state\cite{OP4}, scintillation  \cite{OP5}, structure functions \cite{OP6}, \cite{OP7}, beam wander \cite{OP8}, etc. of light beams generated by sources with a variety of physical and statistical properties have been theoretically predicted.   

The main aim of this paper is to set up the framework for optical beam interaction with the ABCD systems in general and with the bi-static LIDAR systems, in particular, that operate in the presence of the clear-water oceanic turbulence described by power spectrum model in Ref. \cite{ar01}. In contrast with Refs. \cite{AtmLIDAR1}, \cite{AtmLIDAR2} we first derive a simple and elegant transformation law of the beam's second-order correlation function and demonstrate its application to LIDAR links with one or more interactions with targets. Even though the general method is valid for any electromagnetic sources, in numerical calculations we restrict ourselves to the linearly polarized ones. 

The paper is organized as follows: in section 2 the wide-range Prandtl/Schmidt power spectrum model is reviewed; in section 3 the general method is developed for beam propagation through an ABCD optical system placed into the turbulent oceanic column, expressed in terms of a single matrix transformation; in section 4 the general method of section 3 is adopted to bi-static underwater LIDARs with the help of two consecutive matrix transformations; in section 5 numerical results for the second-order beam statistics (the spectral density and the degree of coherence) in the special case of the single pass propagation are analyzed for a variety of average water temperatures; in section 6 the LIDAR system is analyzed for a variety of the turbulence parameters; section 7 summarizes the results. 

\section{The oceanic turbulence power spectrum}
\label{sec:modelspectrum}
We begin by reviewing the power spectrum model of oceanic turbulence developed in Ref. \cite{ar01} that is capable of accounting for the wide ranges of the Prandtl/Schmidt numbers. In general, optical turbulence in an oceanic column is caused by mechanical mixing of water with different temperatures and salinity concentrations. Two separate spatial power spectra of the refractive index fluctuations due to temperature fluctuations, $\Phi_{\mathrm{T}}(\kappa)$, and salinity fluctuations, $\Phi_{\mathrm{S}}(\kappa)$ can be obtained and combined into the single spectrum $\Phi_{\mathrm{n}}(\kappa)$, for instance, by a polynomial linearization procedure \cite{ar01}:
\begin{align}
\Phi_{\mathrm{n}}(\kappa)=A^{2} \Phi_{\mathrm{T}}(\kappa)+B^{2} \Phi_{\mathrm{S}}(\kappa)-2 A B \Phi_{\mathrm{TS}}(\kappa),
\label{eq:spr}
\end{align}
where
\begin{align}
\nonumber\Phi_{i}(\kappa)=&\frac{ \beta \varepsilon^{-\frac{1}{3}} \kappa^{-\frac{11}{3}} \chi_{i} }{4 \pi}\left[1+21.61(\kappa \eta)^{0.61} c_{i}^{0.02}-18.18(\kappa \eta)^{0.55} c_{i}^{0.04}\right]\\
&\times \exp \left[-174.90(\kappa \eta)^{2} c_{i}^{0.96}\right], \qquad\qquad i \in\{\mathrm{T}, \mathrm{S}, \mathrm{TS}\},
\label{eq:eq2}
\end{align}
and ${c_i} = {0.072^{4/3}\beta {Pr }_i^{-1}}$, ${Pr }_{\rm{T}}$ and ${Pr }_{\rm{S}}$ are the temperature Prandtl number and the salinity Schmidt number, respectively, while ${Pr _{\rm{TS}}} = 2{Pr _{\rm{T}}}{Pr _{\rm{S}}}({Pr _{\rm{T}}}+{Pr _{\rm{S}}})^{-1}$ is the coupled Prandtl-Schmidt number. As shown in Ref. \cite{ar01} and Appendix, at different naturally occurring average water temperatures in the interval $[0^{\circ}\rm C, 30^{\circ}\rm C]$, ${Pr}_{\rm{T}}$, ${Pr}_{\rm{S}}$ and ${Pr}_{\rm{TS}}$ may substantially vary leading to noticeable effects on the light statistics. Also, in Eq. (2) $\eta$ is the Kolmogorov microscale, and ${\chi _i}$ are the ensemble-averaged variance dissipation rates which are related by expressions 
\begin{align}
{\chi _{\rm{S}}} = \frac{{{A^2}}}{{{\omega ^2}{B^2}}}{\chi _{\rm{T}}}{d_r}, \quad {\chi _{{\rm{TS}}}} = \frac{A}{{2\omega B}}{\chi _{\rm{T}}}\left( {1 + {d_r}} \right),
\label{eq10}
\end{align}
${d_r}$ being the eddy diffusivity ratio and $\omega $ being the relative strength of temperature-salinity fluctuations.
In most practical cases, 
${d_r}$ and $\omega $ are related as
\begin{align}
{d_r} \approx \left\{ {
	\begin{array}{ll}
	{\left| \omega  \right|} + {\left| \omega  \right|^{0.5}}{\left( {\left| \omega  \right| - 1} \right)^{0.5},} & {\left| \omega  \right| \ge 1}\\
	{1.85\left| \omega  \right| - 0.85,} & {0.5 \le \left| \omega  \right| < 1}\\
	{0.15\left| \omega  \right|,} & {\left| \omega  \right| < 0.5}
	\end{array}}
\right..
\label{eq07_3}
\end{align}

We note that parameters ${Pr}_{\rm{T}}$, ${Pr}_{\rm{S}}$, ${Pr}_{\rm{TS}}$ and $\eta$ vary with average temperature $\left\langle T \right\rangle$, $\varepsilon$ is dominated by velocity field, $\omega$ and $\chi _{\rm{T}}$ are governed by the scalar gradient and the turbulent eddy diffusivity. We will discuss in detail the influence of $\left\langle T \right\rangle$, $\varepsilon$, $\omega$ and $\chi _{\rm{T}}$ on the beam passage in the underwater turbulence in the following sections.

\section{Transformation of the cross-spectral density matrix in an ABCD system}
\label{sec:matrixapp}

In this section, we develop a concise $4\times 4$ ABCD matrix method for predicting the evolution of the Cross-Spectral Density (CSD) of a light beam in complex optical systems embedded in any linear turbulent medium. The evolution of the electric field $\textbf{E}=\{E_x,E_y\}$ from one, input, plane ($z=0$) to another, output, plane ($z=L$) in the presence of an ABCD optical system and the turbulent fluctuations can be found with the help of the extended Huygens-Fresnel integral \cite{AtmLIDAR1}, \cite{AtmLIDAR2}:
\begin{align}
\nonumber{E_\alpha }({\bf{t}},L) =& \frac{{ - ik}}{{2\pi \sqrt {Det[{\bf{B}}]} }}\iint {{E_\alpha }({\bf{s}},0)\exp \left[ {\psi ({\bf{s}},{\bf{t}})} \right]}\\
&\times \exp \left[ {\frac{{ - ik}}{2}({{\bf{s}}^T}{{\bf{B}}^{ - 1}}{\bf{As}} - 2{{\bf{s}}^T}{{\bf{B}}^{ - 1}}{\bf{t}} + {{\bf{t}}^T}{{\bf{D}}^{ - 1}}{\bf{t}})} \right]d{\bf{s}}.
\label{eq:HF_ing}
\end{align}
Here $\textbf{s}=(s_x,s_y)^T$ and $\textbf{t}=(t_x,t_y)^T$ are transverse position vectors of points in the input and output planes, subscript $T$ stands for transpose, $k=2\pi/\lambda$ is the wavenumber, $\lambda$ being the speed of light in vacuum, $\psi(\textbf{s},\textbf{t})$ is the complex phase perturbation of the spherical wave by turbulence, $Det$ stands for matrix determinant and $\textbf{A}$, $\textbf{B}$,  $\textbf{C}$ and  $\textbf{D}$ are the $2\times 2$ submatrices of the astigmatic optical system, corresponding to the scalar ABCD matrix elements, i.e., $\textbf{A}=A \textbf{I}$, $\textbf{B}=B \textbf{I}$, $\textbf{C}=C \textbf{I}$ and $\textbf{D}=D \textbf{I}$, $\textbf{I}$ being the 2D identity matrix, and integration is performed over the entire source plane. On correlating the electric field components in Eqs. (\ref{eq:HF_ing}) we obtain the elements of the CSD matrix:
\begin{align}
\nonumber
&W_{\alpha \beta }^{(T)}({{\bf{t}}_1},{{\bf{t}}_2},L) 
= \langle {E_\alpha }({{\bf{t}}_1},L)E_\beta ^*({{\bf{t}}_2},L)\rangle \\
\nonumber
= &\frac{{{k^2}}}{{4{\pi ^2}\sqrt {Det[{\bf{B}}]Det[{{\bf{B}}^*}]} }} \iint{W_{\alpha \beta }^{(S)}({{\bf{s}}_1},{{\bf{s}}_2},0){\Gamma _\psi }({{\bf{t}}_1},{{\bf{t}}_2},{{\bf{s}}_1},{{\bf{s}}_2})}\\
\nonumber
&\times \exp \left[ { - \frac{{ik}}{2}({\bf{s}}_1^T{\bf{B}}_\alpha ^{ - 1}{{\bf{A}}_\alpha }{{\bf{s}}_1} - 2{\bf{s}}_1^T{\bf{B}}_\alpha ^{ - 1}{{\bf{t}}_1} + {\bf{t}}_1^T{{\bf{D}}_\alpha }{\bf{B}}_\alpha ^{ - 1}{{\bf{t}}_1})} \right]\\
&\times \exp \left[ { \frac{{ik}}{2}({\bf{s}}_2^T{\bf{B}}_\beta ^{ - 1}{{\bf{A}}_\beta }{{\bf{s}}_2} - 2{\bf{s}}_2^T{\bf{B}}_\beta ^{ - 1}{{\bf{t}}_2} + {\bf{t}}_2^T{{\bf{D}}_\beta }{\bf{B}}_\beta ^{ - 1}{{\bf{t}}_2})} \right]d{{\bf{s}}_1}d{{\bf{s}}_2},
\label{eq:SCC}
\end{align}
where
\begin{align}
W_{\alpha\beta}^{(S)}(\textbf{s}_1, \textbf{s}_2, 0)=\langle E_{\alpha}(\textbf{s}_1,0)E_{\beta}^*(\textbf{s}_2,0)  \rangle
\end{align}
describes the CSD of source. Further, $\Gamma _\psi$ is the phase correlation function of the spherical wave \cite{AP}:
\begin{align}
\nonumber &{\Gamma _\psi(\textbf{t}_1, \textbf{t}_2,\textbf{s}_1, \textbf{s}_2)} =\left\langle\exp \left[\psi\left(\mathbf{s}_{1}, \mathbf{t}_{1}\right)+\psi^{*}\left(\mathbf{s}_{2}, \mathbf{t}_{2}\right)\right]\right\rangle \\
& =\exp \left[-\frac{1}{\rho_{0}^{2}}\left[\left(\mathbf{s}_{1}-\mathbf{s}_{2}\right)^{2}+\left(\mathbf{s}_{1}-\mathbf{s}_{2}\right)\left(\mathbf{t}_{1}-\mathbf{t}_{2}\right)+\left(\mathbf{t}_{1}-\mathbf{t}_{2}\right)^{2}\right]\right],
\label{eq:PCF}
\end{align}
with $\rho_{0}$ [m] being the coherence radius, which is determined by setting $\Gamma _\psi$  to value 2 for fixed propagation distance $z=L$:
\begin{align}
D_{s p}\left(\rho_{0}, L\right)=2.
\label{eq:eqforDsp}
\end{align}
Within the validity of the Rytov approximation such structure function has the form \cite{AP}
\begin{align}\label{eq:D}
{D_{{\rm{sp}}}}(\rho ,L) &= 8{\pi ^2}{k^2}L\int_0^1 {\int_0^\infty  \kappa  } {\Phi _n}(\kappa )\left[ {1 - {J_0}(\kappa \xi \rho )} \right]d\kappa d\xi,
\end{align}
where $J_0$ is the Bessel function of the first kind and order zero.

Next, on introducing the 4D vectors $\tilde{\textbf{s}}=(s_{1x},s_{1y},s_{2x},s_{2y})^T$ and $\tilde{\textbf{t}}=(t_{1x},t_{1y},t_{2x},t_{2y})^T$ for the input and output planes, respectively, as well as $4\times 4$ matrices 
\begin{equation}
\begin{array}{*{20}{c}}
{\tilde {\bf{A}} = \left[ {\begin{array}{*{20}{c}}
		{{{\bf{A}}_\alpha }}&{0{\bf{I}}}\\
		{0{\bf{I}}}&{{\bf{A}}_\beta ^*}
		\end{array}} \right],\quad \tilde {\bf{B}} = \left[ {\begin{array}{*{20}{c}}
		{{{\bf{B}}_\alpha }}&{0{\bf{I}}}\\
		{0{\bf{I}}}&{ - {\bf{B}}_\beta ^*}
		\end{array}} \right],\quad \tilde {\bf{C}} = \left[ {\begin{array}{*{20}{c}}
		{{{\bf{C}}_\alpha }}&{0{\bf{I}}}\\
		{0{\bf{I}}}&{ - {\bf{C}}_\beta ^*}
		\end{array}} \right],}\\
{\tilde {\bf{D}} = \left[ {\begin{array}{*{20}{c}}
		{{{\bf{D}}_\alpha }}&{0{\bf{I}}}\\
		{0{\bf{I}}}&{{\bf{D}}_\beta ^*}
		\end{array}} \right],\quad \tilde {\bf{O}} = \frac{2}{{ik\rho _0^2}}\left[ {\begin{array}{*{20}{c}}
		{\bf{I}}&{ - {\bf{I}}}\\
		{ - {\bf{I}}}&{\bf{I}}
		\end{array}} \right],}
\end{array}
\end{equation} 
where $\textbf{I}$ is the $2\times 2$ identity matrix, and $\rho_{0}$ is given in Eq. \eqref{eq:eqforDsp}. In view of these definitions, on substituting from  Eq. \eqref{eq:PCF} into Eq. \eqref{eq:SCC} we get the concise propagation law
\begin{align}
\nonumber 
{W_{\alpha \beta }}(\tilde {\bf{t}},L) &= \frac{{{k^2}}}{{4{\pi ^2}\sqrt {Det[\tilde {\bf{B}}]} }}\iint {W_{\alpha \beta }^{(S)}(\tilde {\bf{s}},0)} \\
\nonumber 
&\times \exp \left[ { - \frac{{ik}}{2}\left( {{{\tilde {\bf{s}}}^T}[{{\tilde {\bf{B}}}^{ - 1}}\tilde {\bf{A}} + \tilde {\bf{O}}]\tilde {\bf{s}} + {{\tilde {\bf{t}}}^T}[\tilde {\bf{D}}{{\tilde {\bf{B}}}^{ - 1}} + \tilde {\bf{O}}]\tilde {\bf{t}}} \right)} \right]
\\ 
&\times \exp \left[ {\frac{{ik}}{2}\left( {2{{\tilde {\bf{s}}}^T}[{{\tilde {\bf{B}}}^{ - 1}} - \tilde {\bf{O}}/2]\tilde {\bf{t}}} \right)} \right]d\tilde {\bf{s}}.
\label{eq:W2}
\end{align}

We now limit ourselves to the electromagnetic Gaussian Schell-model (GSM) source \cite{KorBook}:
\begin{align}
W_{\alpha\beta}^{(S)}(\tilde{\textbf{s}},0)=I_{\alpha\beta}^{(s)} \exp\left[ -\frac{ik}{2}\tilde{\textbf{s}}^T \textbf{S}^{-1}_{\alpha\beta}\tilde{\textbf{s}}  \right], \quad (\alpha,\beta=x,y),
\label{eq:Ws}
\end{align}
where
\begin{align}
I_{\alpha\beta}^{(s)}=A_{\alpha}A_{\beta}B_{\alpha\beta}, \quad 
\textbf{S}^{-1}_{\alpha\beta}= \begin{bmatrix}
\frac{1}{ik}\left(\frac{1}{2\sigma_{\alpha}^2}+\frac{1}{\delta_{\alpha\beta}^2} \right)\textbf{I} & \frac{i}{k\delta_{\alpha\beta}^2}\textbf{I}  \\
\frac{i}{k\delta_{\alpha\beta}^2}\textbf{I}  &  \frac{1}{ik}\left(\frac{1}{2\sigma_{\beta}^2}+\frac{1}{\delta_{\alpha\beta}^2} \right)\textbf{I}
\end{bmatrix}.
\label{eq:M0}
\end{align} 
Here $\delta_{\alpha\beta}$ are the root-mean-square (rms) correlation widths and $\sigma_{\alpha}$ are the rms spectral density widths, $A_{\alpha}$ are the amplitudes, and $B_{xy}$ is the the maximum value of correlation between $x$ and $y$ electric field components ($B_{xx}=B_{yy}=1$). The parameters have to satisfy realizability conditions. This model can be used to assign arbitrary spectral, coherence, polarization and cross-polarization properties of electromagnetic beam-like beams that can change on propagation. On substituting from Eqs. \eqref{eq:Ws} and \eqref{eq:M0} into Eq. \eqref{eq:W2} we arrive at the expression
\begin{align}
\nonumber
{W_{\alpha \beta }}(\tilde {\bf{t}},L) = &\iint {\exp \left[ { - \frac{{ik}}{2}\left( {{{\tilde {\bf{s}}}^T}{{\tilde {\bf{X}}}_{1\alpha \beta }}\tilde {\bf{s}} - 2{{\tilde {\bf{s}}}^T}{{\tilde {\bf{X}}}_2}\tilde {\bf{t}} + {{\tilde {\bf{t}}}^T}{{\tilde {\bf{X}}}_3}\tilde {\bf{t}}} \right)} \right]d\tilde {\bf{s}}} \\
&\times \frac{{I_{\alpha \beta }^{(s)}{k^2}}}{{4{\pi ^2}\sqrt {Det[\tilde {\bf{B}}]} }},
\end{align}
where $4\times 4$ matrices have forms
\begin{align}
\begin{array}{*{20}{c}}
\tilde{\textbf{X}}_{1\alpha\beta}=\textbf{S}^{-1}_{\alpha\beta}+\tilde{\textbf{B}}^{-1}\tilde{\textbf{A}}+\tilde{\textbf{O}},\\
\tilde{\textbf{X}}_2=\tilde{\textbf{B}}^{-1}-\tilde{\textbf{O}}/2, \quad \tilde{\textbf{X}}_3=\tilde{\textbf{D}}\tilde{\textbf{B}}^{-1}+
\tilde{\textbf{O}}.
\end{array}
\end{align}

It follows from the matrix identity 
\begin{equation}
\tilde{\textbf{s}}^T\tilde{\textbf{X}}_{1\alpha\beta}\tilde{\textbf{s}}
-2\tilde{\textbf{s}}^T\tilde{\textbf{X}}_2\tilde{\textbf{t}} =\tilde{\textbf{t}}^T\tilde{\textbf{X}}_2^T \tilde{\textbf{X}}_{1\alpha\beta}^{-1}\tilde{\textbf{X}}_2\tilde{\textbf{t}}+|\tilde{\textbf{X}}_{1\alpha\beta}^{1/2}\tilde{\textbf{s}}-\tilde{\textbf{X}}_{1\alpha\beta}^{-1/2}\tilde{\textbf{X}}_2\tilde{\textbf{t}}|^2
\end{equation}
and the Gaussian integral formula
\begin{equation}
\int\limits_{-\infty}^{\infty}\exp[-ax^2]=\sqrt{\frac{\pi}{a}}
\end{equation}
that 
\begin{equation}
W_{\alpha\beta}(\tilde{\textbf{t}},L)=I^{(Y)}_{\alpha\beta}\exp \left[-\frac{ik}{2}
\tilde{\textbf{t}}^T\tilde{\textbf{Y}}_{\alpha\beta}\tilde{\textbf{t}}\right],
\label{eq:WatT}
\end{equation}
where
\begin{equation}
I^{(Y)}_{\alpha\beta}=\frac{I^{(s)}_{\alpha\beta}}{ \sqrt{Det[\tilde{\textbf{B}}]Det[\tilde{\textbf{X}}_{1\alpha\beta}]}}, \quad \tilde{\textbf{Y}}_{\alpha\beta}=\tilde{\textbf{X}}_3-\tilde{\textbf{X}}_2^T \tilde{\textbf{X}}_{1\alpha\beta}^{-1}\tilde{\textbf{X}}_2.
\label{eq:eq20}
\end{equation} 
Formula \eqref{eq:WatT} is the fundamental transformation of the CSD matrix of the optical field in the input plane to the field in the output plane. Thus, provided the source is of the electromagnetic Gaussian Schell type and the quadratic approximation to the spherical-wave structure function is made, the CSD remains shape-invariant.

\section{Theoretical framework of Bi-static LIDAR systems}
Since a bi-static system does not involve correlations between the incident and the reflected waves,  light evolution through it can be examined sequentially, the passage from source to target as step one, the interaction with the target as step two, and the passage from target to detecting system as step three. In general, for two propagation paths in turbulence $\rho_0$ must be evaluated separately, according to the formulas in section \ref{sec:modelspectrum}.

\begin{figure*}
	\centering
	\includegraphics[width=0.85\textwidth]{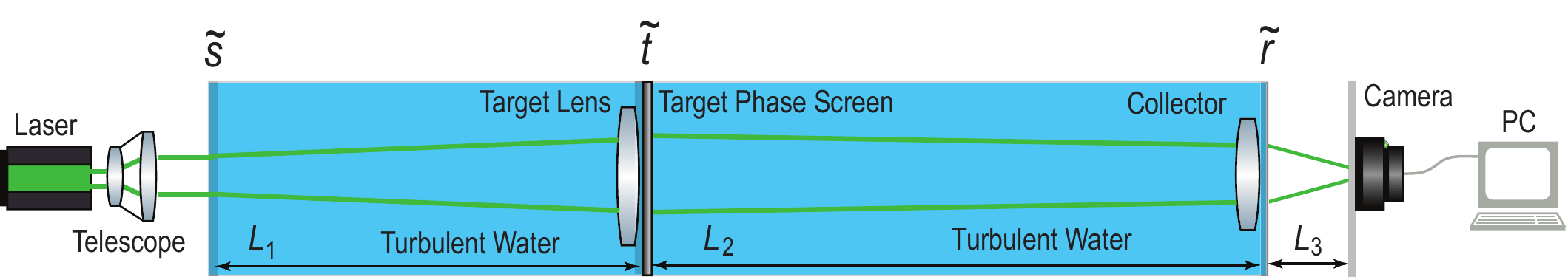}
	\caption{Bi-static underwater LIDAR system (unfolded).}
	\label{fg:f2}
\end{figure*}

Figure \ref{fg:f2} presents a schematic diagram of a generic (unfolded) bi-static LIDAR system of range $L_1$ from the source to the target, and range $L_2$ from the target to the receiver. The target may be modeled as a combination of a Goodman-like thin random phase screen and a Gaussian lens of focal distance $f_1$. After propagation at total distance $L_1 +L_2$ the light beam can be collected by a lens with focal length $f_2$ and focused into a detector, say at distance $L_3$ from it.

According to Fig. \ref{fg:f2} the optical system including propagation at distance $L_1$ and a Gaussian lens with focal distance $f_1$ results in the $4 \times 4 \ \rm{ABCD}$ matrix of the form
\begin{align}
\nonumber\begin{bmatrix} \textbf{A} & \textbf{B} \\ \textbf{C} & \textbf{D} \end{bmatrix}&=
\begin{bmatrix} \textbf{I} & 0\textbf{I} \\ (-1/f_1)\textbf{I} & \textbf{I} \end{bmatrix} \begin{bmatrix} \textbf{I} & L_1\textbf{I} \\ 0\textbf{I} & \textbf{I} \end{bmatrix}\\
&=
\begin{bmatrix} \textbf{I} & L_1\textbf{I} \\ (-1/f_1)\textbf{I} & (1-l_1/f_1)\textbf{I} \end{bmatrix}.
\end{align}
Also, the back-propagation matrix from the target plane to any plane $z>0$ takes form
\begin{align}
\nonumber\begin{bmatrix} \textbf{A}(z) & \textbf{B}(z) \\ \textbf{C}(z) & \textbf{D}(z) \end{bmatrix}&=
\begin{bmatrix} 1 & L_1-z \\ 0 & 1 \end{bmatrix} \begin{bmatrix} 1& 0 \\ -1/f_1 & 1 \end{bmatrix}\\
&=
\begin{bmatrix} 1+(z-L_1)/f_1 & L_1-z  \\ (-1/f_1) & 1 \end{bmatrix}.
\end{align}

The CSD matrix just after interacting with the target lens and just before passing through the random phase screen can then be found from Eq. \eqref{eq:WatT} with $L = L_1$.

Let us now assume that the beam with the CSD matrix in Eq. \eqref{eq:WatT} illuminates a target with spatial correlation function
\begin{align}
C_{T}(\tilde{\mathbf{t}})=I_{T} \exp \left[-\frac{i k}{2} \tilde{\mathbf{t}}^{T} \tilde{\mathbf{T}} \tilde{\mathbf{t}}\right]
\end{align}
where $\tilde{\mathbf{t}}$ now refers to a vector in the target plane; $I_t = 1$ for smooth targets but becomes proportional to the variance of the target reflection coefficient for semi-rough and Lambertian targets and
\begin{align}
\tilde{\textbf{T}}= 2\begin{bmatrix}
\frac{1}{ik}\left(\frac{1}{\sigma_T^2}+\frac{1}{\delta_T^2} \right)\textbf{I} & \frac{i}{k\delta_T^2}\textbf{I}  \\
\frac{i}{k\delta_T^2}\textbf{I}  &  \frac{1}{ik}\left(\frac{1}{\sigma_T^2}+\frac{1}{\delta_T^2} \right)\textbf{I} 
\end{bmatrix} ,
\end{align}
involving soft target size $\sigma_T$ tending to zero and infinity for point and infinite targets and the typical transverse correlation width $\delta_T$, tending to zero and infinity for Lambertian and smooth targets, respectively. 
We assume that the interaction of the illumination with the target surface is local and, hence, multiplicative, i.e., right after interaction with the random phase screen the CSD of the beam becomes
\begin{align}
\nonumber W^{(T)}_{\alpha\beta}(\tilde{\textbf{t}},L_1)&=W_{\alpha\beta}(\tilde{\textbf{t}},L_1) C_T(\tilde{\textbf{t}})  \\&
=I^{(Z)}_{\alpha\beta}\exp \left[-\frac{ik}{2}
\tilde{\textbf{t}}^T\tilde{\textbf{Z}}_{\alpha\beta}\tilde{\textbf{t}}\right], 
\label{Wz1}
\end{align}
where superscript $(T)$ is used to distinguish CSD from the one in Eq. \eqref{eq:WatT}, and 
\begin{align}
I^{(Z)}_{\alpha\beta}=I^{(Y)}_{\alpha\beta}I_T, \quad \tilde{\textbf{Z}}_{\alpha\beta}=\tilde{\textbf{Y}}_{\alpha\beta}+\tilde{\textbf{T}}.
\label{eq:eq27}
\end{align}

Thus, the total transformation from  $W^{(S)}_{\alpha\beta}(\tilde{s},0)$ to $W^{(T)}_{\alpha\beta}(\tilde{t},L_1)$ based on the Schell-model of the source and of the target surface as well as quadratic approximation for the spherical wave structure function is shape-invariant as well. The CSD function in Eq. \eqref{Wz1}  can be now regarded as a secondary, planar random source for further passage through the turbulent channel. For instance, if at distance $L_2$ from plane $z=L_1$ the beam is received by a Gaussian collecting lens with focal length $f_2$ (see Fig. 2) then right after the lens the CSD of the beam specified at the $4\times 4$ vector $\tilde{\textbf{r}} =(r_{1x},r_{1y},r_{2x},r_{2y})^T$ can be described by another ``Y''-type transformation, denoted by a prime:  
\begin{align}\label{WatR}
W'_{\alpha\beta}(\tilde{\textbf{r}},L_1+L_2)=I^{(Y')}_{\alpha\beta}\exp \left[-\frac{ik}{2}
\tilde{\textbf{t}}^T\tilde{\textbf{Y}}'_{\alpha\beta}\tilde{\textbf{t}}\right]
\end{align}
where
\begin{equation}
I^{(Y')}_{\alpha\beta}=\frac{I^{(Z)}_{\alpha\beta}}{ \sqrt{Det[\tilde{\textbf{B}'}]Det[\tilde{\textbf{X}}'_{1\alpha\beta}]}}, \quad \tilde{\textbf{Y}}'_{\alpha\beta}=\tilde{\textbf{X}}'_3-\tilde{\textbf{X}'}_2^T \tilde{\textbf{X}'}_{1\alpha\beta}^{-1}\tilde{\textbf{X}'}_2,
\label{eq:eq29}
\end{equation} 
and
\begin{align}
\begin{array}{*{20}{c}}
\tilde{\textbf{X}}'_{1\alpha\beta}=\textbf{Z}_{\alpha\beta}+\tilde{\textbf{B}'}^{-1}\tilde{\textbf{A}}'
+\tilde{\textbf{O}}',\\
\tilde{\textbf{X}}'_2=\tilde{\textbf{B}}'^{-1}-\tilde{\textbf{O}}'/2, \quad \tilde{\textbf{X}}'_3=\tilde{\textbf{D}}'\tilde{\textbf{B}'}^{-1}+
\tilde{\textbf{O}}'.
\end{array}
\end{align}
In these formulas all the ABCD matrix elements with a prime, including those of $\tilde{\textbf{O}}'$ (depending on a new spherical wave coherence radius $\rho'_0$) are obtained by simple replacement of $L_1$ and $f_1$ by $L_2$ and $f_2$. Also if a turbulent channel has another spectrum, say $\Phi'_n(\kappa)$ then it must be used instead of $\Phi_n(\kappa)$ in calculations of $\rho_0'$.

Along the same lines, if at distance $L_2$ from plane $z=L_1$ there is a second target with the surface correlation function $C_{T'}(\tilde{r})$ (not shown on Fig. 2) then another ``Z''-type transformation, denoted by Z', can be applied:
\begin{align}
\nonumber W^{(T')}_{\alpha\beta}(\tilde{\textbf{r}},L_1+L_2)&=W'_{\alpha\beta}(\tilde{\textbf{r}},L_1+L_2) C_{T'}(\tilde{\textbf{r}})  \\&
=I^{(Z')}_{\alpha\beta}\exp \left[-\frac{ik}{2}
\tilde{\textbf{r}}^T\tilde{\textbf{Z}}'_{\alpha\beta}\tilde{\textbf{r}}\right], 
\label{Wz}
\end{align}
where 
\begin{align}
\begin{array}{*{20}{c}}
C_{T'}(\tilde{\textbf{r}})= I'_T\exp \left[-\frac{ik}{2}
\tilde{\textbf{t}}^T\tilde{\textbf{T}'}\tilde{\textbf{t}}\right], \quad \\
\\
\tilde{\textbf{T}'}= 2\begin{bmatrix}
\frac{1}{ik}\left(\frac{1}{\sigma_T^{'2}}+\frac{1}{\delta_T^{'2}} \right)\textbf{I} & \frac{i}{k\delta_T^{'2}}\textbf{I}  \\
\frac{i}{k\delta_T^{'2}}\textbf{I}  &  \frac{1}{ik}\left(\frac{1}{\sigma_T^{'2}}+\frac{1}{\delta_T^{'2}} \right)\textbf{I},   
\end{bmatrix}, 
\end{array}
\end{align}
and
\begin{align}
I^{(Z')}_{\alpha\beta}=I^{(Y')}_{\alpha\beta}I_{T'}, \quad \tilde{\textbf{Z}}'_{\alpha\beta}=\tilde{\textbf{Y}}'_{\alpha\beta}+\tilde{\textbf{T}}',
\end{align}
with $I_{T}'$ and $\tilde{\textbf{T}'}$ being the maximum strength and the $4\times 4$ correlation matrix of the second target. Such iterative procedure can be, in principle, used for a number of reflections from targets and subsequent propagation through turbulence.

\begin{figure*}
	\centering
	\includegraphics[width=0.85\textwidth]{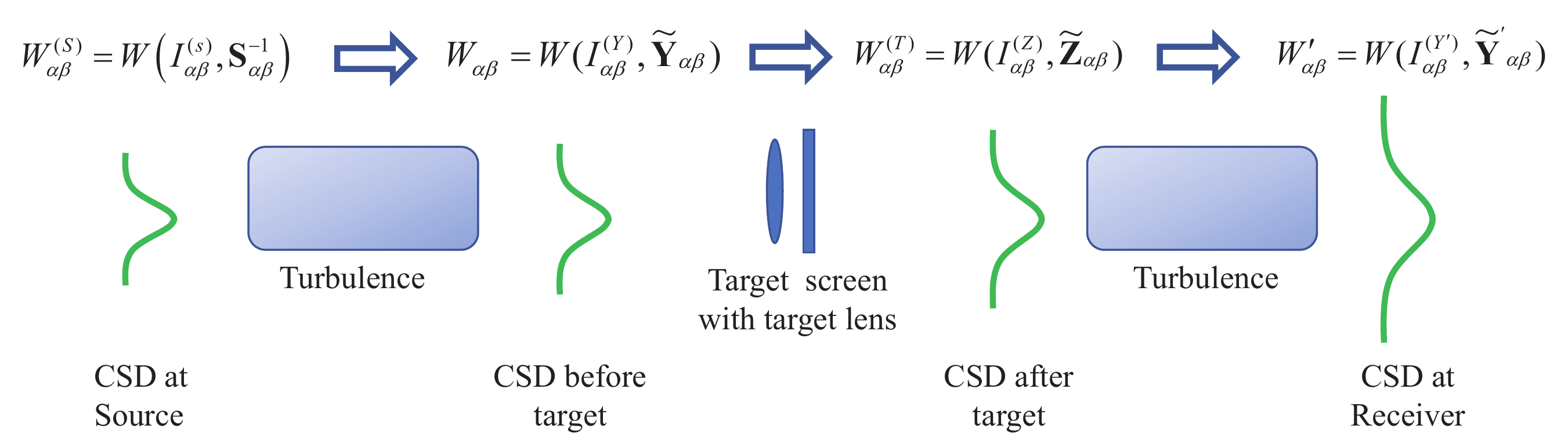}
	\caption{The theoretical framework of Bi-static Lidar.}
	\label{fg:f3}
\end{figure*}

We summarize the evolution of the CSD through the LIDAR system in Fig. \ref{fg:f3}. The CSD propagating through each stage of the LIDAR system is expressed by function $W(I,\tilde{\textbf{Y}}) = I\exp [ - ik{\tilde{\textbf{n}}^{\rm{T}}}\tilde{\textbf{Y}}\tilde{\textbf{n}}/2]$. At each stage $I$ and $\tilde{\textbf{Y}}$ can be easily obtained from Eqs. \eqref{eq:eq20}, \eqref{eq:eq27} and \eqref{eq:eq29}. 

\section{Second-order statistics of optical beams in oceanic turbulence with different average temperatures}

Before considering numerical examples relating to beam passage through the bi-static LIDAR systems we first analyze the evolution of the second-order beam statistics along the single path (in the absence of any optical elements). In this case the ABCD matrix elements reduce to:
\begin{align}
\begin{array}{*{20}{c}}
\textbf{A}_x = \textbf{A}_y = 1 \textbf{I},\quad
\textbf{B}_x = \textbf{B}_y = L \textbf{I},\\
\textbf{C}_x = \textbf{C}_y = 0 \textbf{I},\quad
\textbf{D}_x = \textbf{D}_y = 1 \textbf{I}.
\end{array}
\end{align}

\begin{figure*}
	\centering
	\subfigure{
		\includegraphics[width=0.45\textwidth]{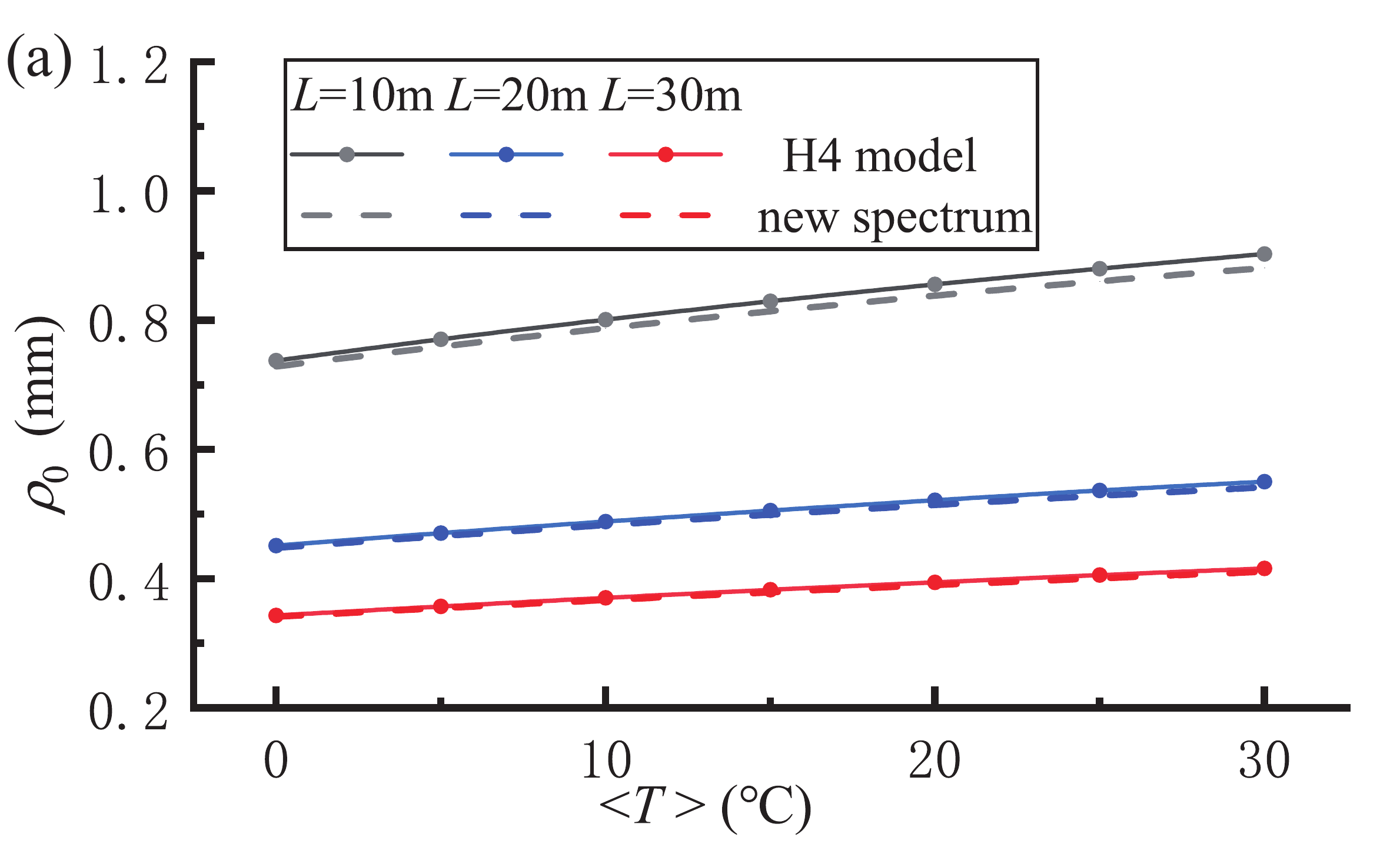}
		\label{fg:f4_a}}
	\subfigure{
		\includegraphics[width=0.45\textwidth]{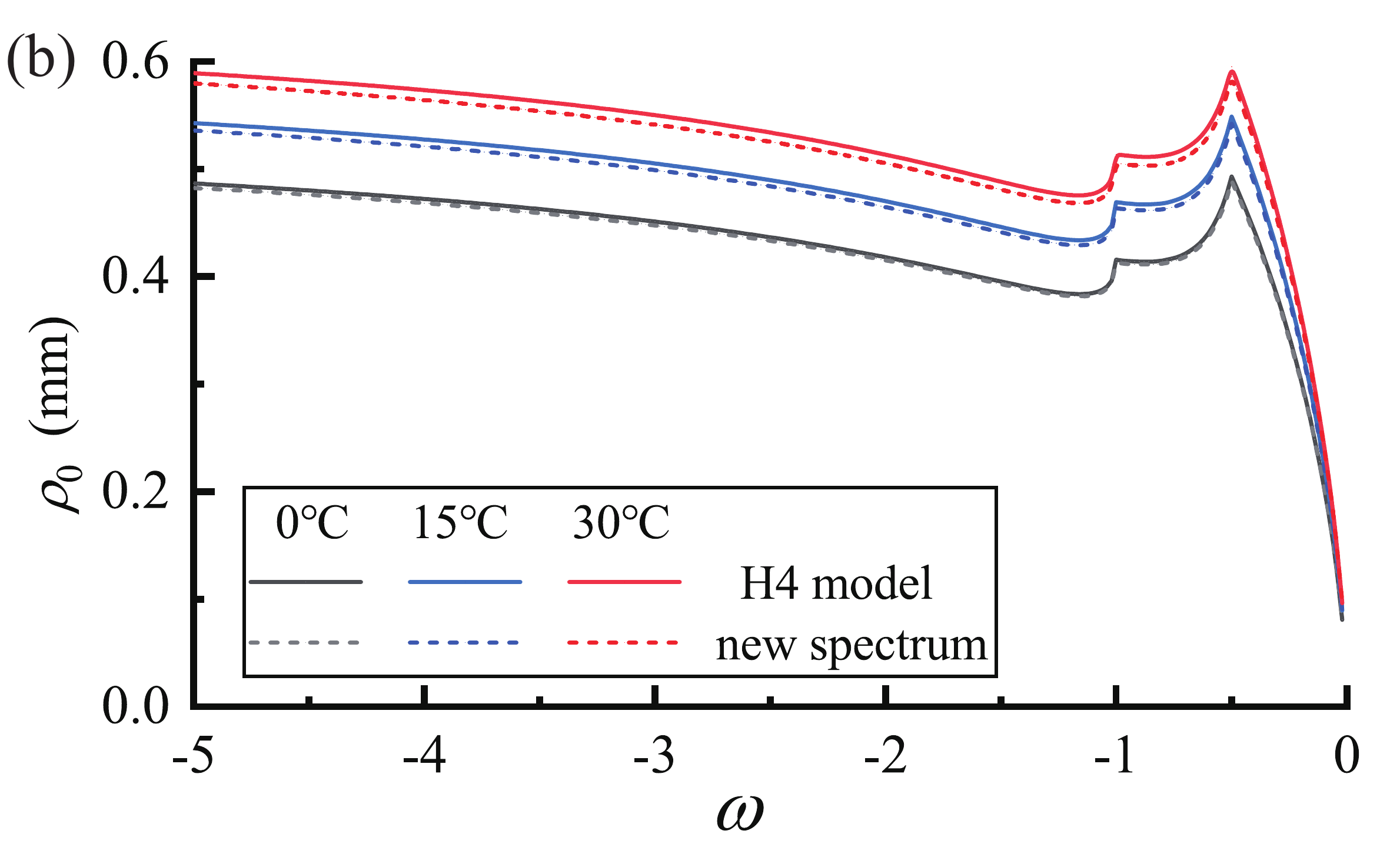}
		\label{fg:f4_b}}
	\caption{Spherical wave coherence radius $\rho_0$ varies with $\left\langle T \right\rangle$, $L$ and $\omega$ in the new spectrum \cite{ar01} and the original Hill's model 4 \cite{Hill2}.}
	\label{fg:f4}
\end{figure*}

First, on using the power spectrum (\ref{eq:spr}) in expression (\ref{eq:D}) for the spherical wave structure function and then solving Eq. (\ref{eq:eqforDsp}) we numerically obtain the coherence radius $\rho_0(L)$ at different propagation distances $L$ and present it in Fig. \ref{fg:f4_a}  for different average temperatures, we also show $\rho_0(\omega)$ varies with $\omega$ in Fig. \ref{fg:f4_b} for different average temperatures $\left\langle T \right\rangle$. The spherical wave is a special case of the electromagnetic GSM beam with $\sigma_i \rightarrow 0$, and $\delta_{ij} \rightarrow \infty$. The figure includes the values of $\rho_0$ as obtained from the analytical fit (\ref{eq:spr}) with numerical solution of the differential equation given by in the original Hill model 4 \cite{Hill2}, which are in a very good agreement. More importantly, we observe that with increase in the  average water temperature the coherence radius increases as well. The dependence of the coherence radius on other parameters have been previously explored (see review \cite{KorOceanRev}). As we now show, this dependence is carried over to the second-order statistics of the finite beams.

The two second-order statistical properties that will be analyzed here are the spectral density $S$ and the degree of coherence $\mu$ are defined by expressions \cite{KorBook}:
\begin{align}
\nonumber &S(\textbf{t},L)=Tr [\textbf{W}(\textbf{t},\textbf{t},L)],\\
&\mu(\textbf{t}_1,\textbf{t}_2,L)=\frac{Tr[ \textbf{W}(\textbf{t}_1,\textbf{t}_2,L)]}{\sqrt{S(\textbf{t}_1,L)}\sqrt{S(\textbf{t}_2,L)}},
\end{align} 
where $Tr$ stands for the matrix trace. Under the assumption that the light beam is linearly polarized along $x$ direction and that the coherence is calculated at two points on $x$-axis, symmetric with respect to the origin, we get $\tilde{\textbf{t}}=(t,0,-t,0)^T$ and these expressions reduce to:
\begin{align}
\nonumber&S(t,L)=W_{xx}((t,0,t,0)^T,L),\\
&\mu(t,L)=\frac{W_{xx}((t,0,-t,0)^T,L)}{W_{xx}((t,0,t,0)^T,L)},
\end{align}
where $W_{xx}$ is calculated from Eq.(\ref{eq:WatT}).

\begin{figure*}
	\centering
	\subfigure{
		\includegraphics[width=0.35\textwidth]{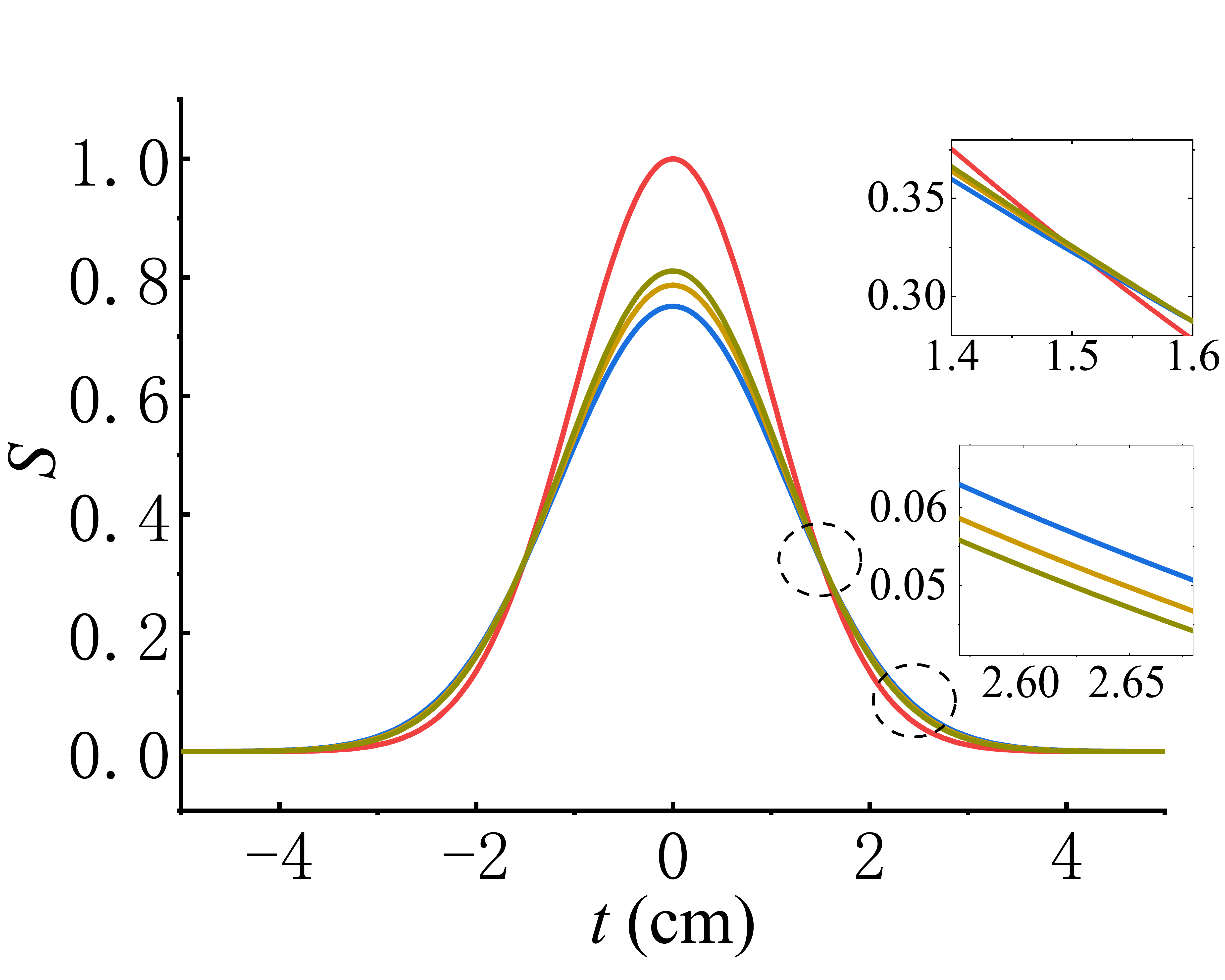}
		\label{fg:f5a}}
	\subfigure{
		\includegraphics[width=0.13\textwidth]{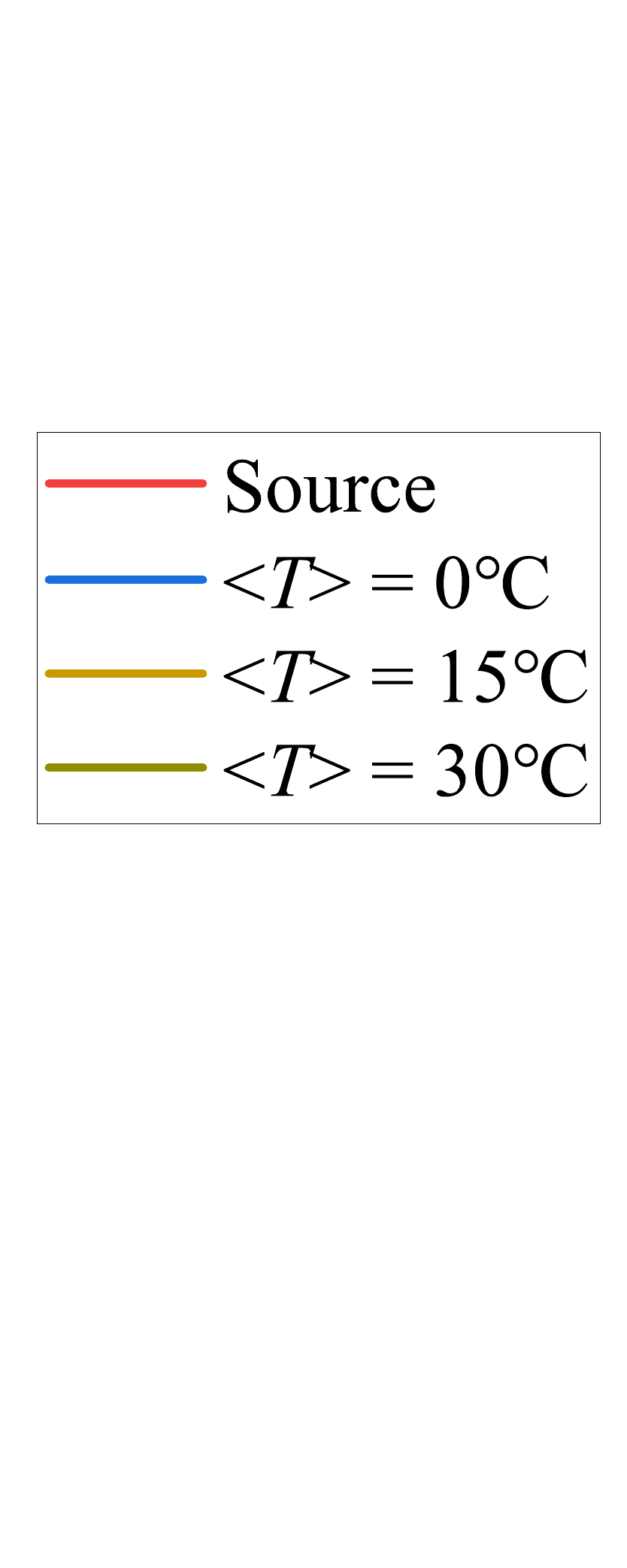}
	}
	\subfigure{
		\includegraphics[width=0.35\textwidth]{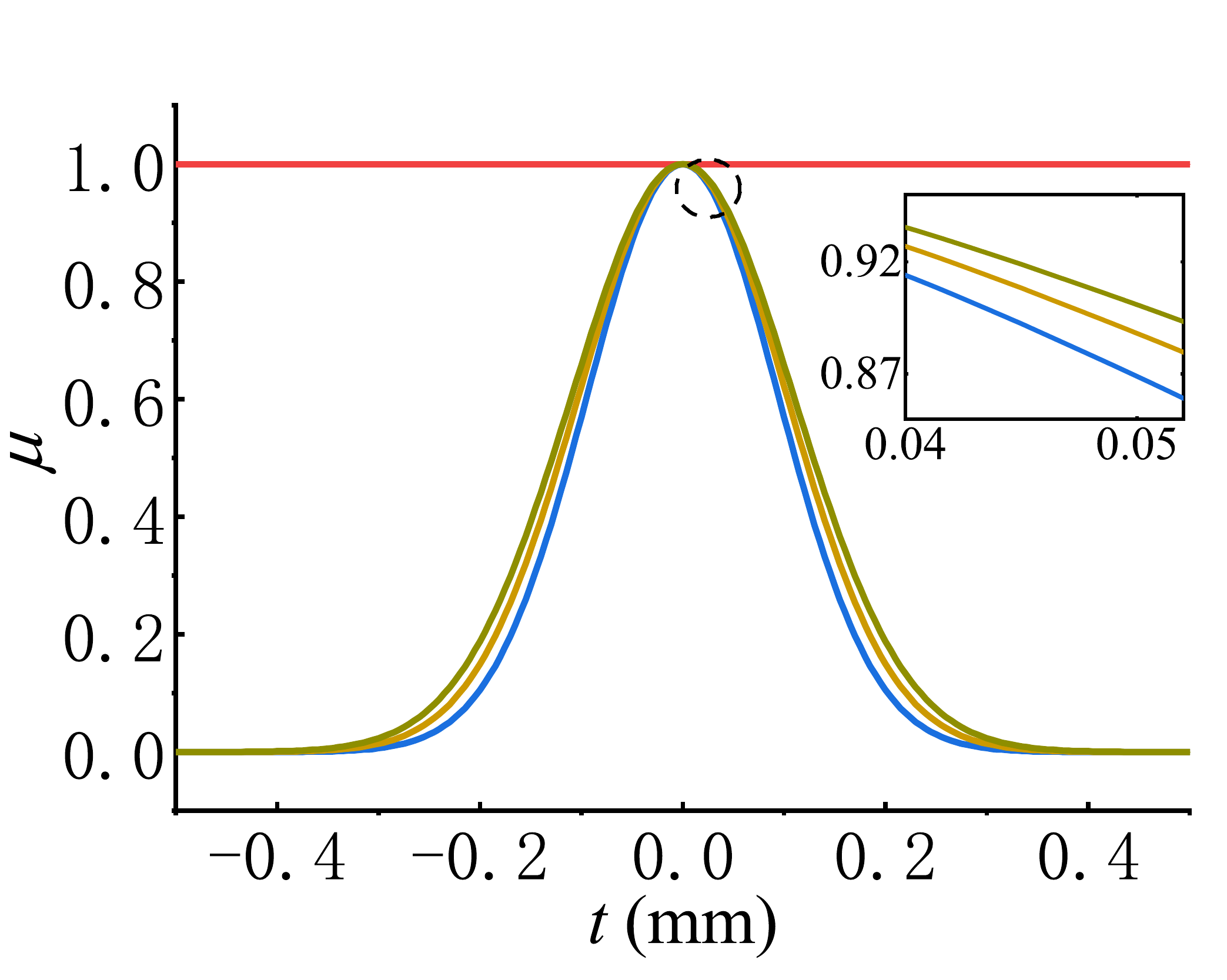}
		\label{fg:f5b}}
	\caption{
		Evolution of $S$ and $\mu$ for different $\left< T \right>$, with $\omega = -3$, $\varepsilon = 10^{-4} \rm m ^2 \rm s ^{-3}$, $\chi_{\rm{T}} = 10^{-5}\rm K ^2 \rm s ^{-1}$, and $L =20 \rm{m}$.
	}
	\label{fg:f5}
\end{figure*}

We assume the parameters of the source source as $\delta_{xx} = 50\rm{cm}$, $\sigma=1 cm$ and  those of the oceanic turbulence as $\omega = -3$, $\varepsilon = 10^{-4} \rm m ^2 \rm s ^{-3}$ and $\chi_{\rm{T}} = 10^{-5} \rm K ^2 \rm s ^{-1}$. The Prandtl number $Pr$ and the Schmidt number $Sc$ vary with average temperature $\left\langle T \right\rangle$, and affect the coherence radius $\rho_{0}$ in Eq.(\ref{eq:eqforDsp}) (see Fig. \ref{fg:f4}), finally leads to the second-order statistics varying with $\left\langle T \right\rangle$  (see Figs. \ref{fg:f5}). Figs. \ref{fg:f4} and \ref{fg:f5} show that a larger average temperature leads to a weaker effects by turbulence, giving a larger coherence radius $\rho_0$ and, hence, keeping a better quality of laser beam, i.e., a lower broadening of intensity $S(t)$ and a smaller decrease in degree of coherence $\mu(t)$. We explain this phenomenon as follows. There are several parameters affect the oceanic turbulence. The kinetic energy dissipation $\varepsilon$ is combined with the velocity field, the temperature (or salinity) dissipation ${\chi _{\rm{T}}}$(or ${\chi _{\rm{S}}}$) is determined by the eddy diffusivity $K_{\rm{T}}$ (or $K_{\rm{S}}$) and the gradient of temperature (or salinity), the Kolmogorov microscale $\eta = \nu ^{3/4}\varepsilon ^{-1/4}$, Prandtl number $Pr _{\rm{T}} = \nu/\alpha_{\rm{T}}$, Schmidt number $Pr _{\rm{S}} = \nu/\alpha_{\rm{S}}$. $\nu$ is the kinematic viscosity, $\alpha_{\rm{S}}$ is the mass diffusivity, $\alpha_{\rm{T}}$ is the thermal diffusivity. \emph{A larger average temperature $\left\langle T \right\rangle$ gives larger $\alpha_{\rm{T}}$, $\alpha_{\rm{S}}$ and a lower $\nu$, thus leads to lower $Pr _{\rm{T}}$, $Pr _{\rm{S}}$ and $\eta$, finally implies a weaker turbulence.} 

\section{The effect of oceanic turbulence on bi-static LIDAR systems}

In this section, we discuss the effect of oceanic turbulence on Bi-static Lidar systems by calculating the evolution of spectral density $S$ and degree of coherence $\mu$ for 
$n = \{s,t,r\}$, i.e., 
\begin{align}
\nonumber &S(n,L)=W_{xx}((n,0,n,0)^T,L), \\
&\mu(n,L)=\frac{W_{xx}((n,0,-n,0)^T,L)}{W_{xx}((n,0,n,0)^T,L)},
\end{align}
including the cases before and after interaction with target for $\tilde{\textbf{t}}$ and the case at receiver for $\tilde{\textbf{r}}$.

As described in Sec.\ref{sec:modelspectrum}, there are several parameters modeling the oceanic turbulence, like rate of dissipation $\chi$, Prandtl numbers $Pr$, Kolmogorov microscale $\eta$ and kinetic energy dissipation $\varepsilon$. 
Among them, the relation between temperature dissipation $\chi_{\rm{T}}$, salinity dissipation $\chi_{\rm{S}}$ and coupled dissipation $\chi_{\rm{TS}}$ is regulated by $\omega $;
$Pr$ and $\eta$ vary a lot with the averaged temperature $\left\langle T \right\rangle$\cite{ar01};
$\varepsilon$ is a statistical parameter of turbulent velocity.
Here we choose $\left\langle T \right\rangle$, $\varepsilon$, $\chi_{\rm{T}}$ and $\omega$ as environment variables, and discuss how they affect the evolution of $S$ and $\mu$.

We set the the source parameters at $\delta_{xx} = 50\,\rm{cm}$ and $\sigma=1 cm$, $\lambda=533nm$, $f=\infty$, and $f=\infty$, $L_1=L_2=20m$. We also choose to set a rough target with a limited size: $I_{T}=0.6$, $\sigma_T = 20\,\rm{cm}$ and $\delta_T = 10\,\rm{cm}$ that

\begin{align}
C_{T}(\tilde{\mathbf{t}})=0.6 \exp \left[-i k \tilde{\mathbf{t}}^{T} \tilde{\mathbf{T}} \tilde{\mathbf{t}}\right],
\end{align}
where
\begin{align}
\tilde{\textbf{T}}= \begin{bmatrix}
-125ik^{-1}\textbf{I} & 100ik^{-1}\textbf{I}  \\
100ik^{-1}\textbf{I}  &  -125ik^{-1}\textbf{I} 
\end{bmatrix}.
\end{align} 

\begin{figure*}
	\centering
	\subfigure{
		\includegraphics[width=0.45\textwidth]{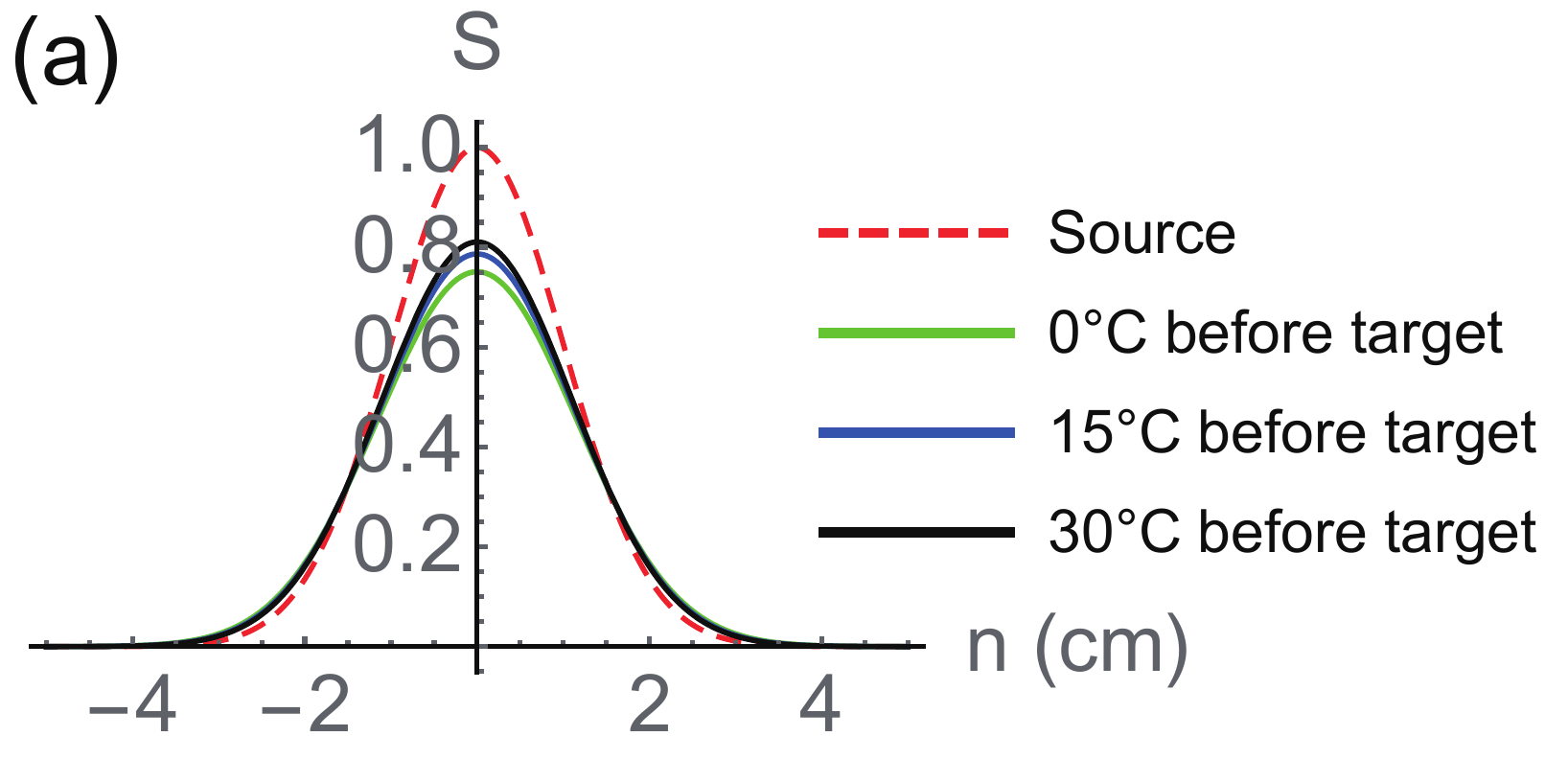}
		\label{fg:f6a}}
	\subfigure{
		\includegraphics[width=0.45\textwidth]{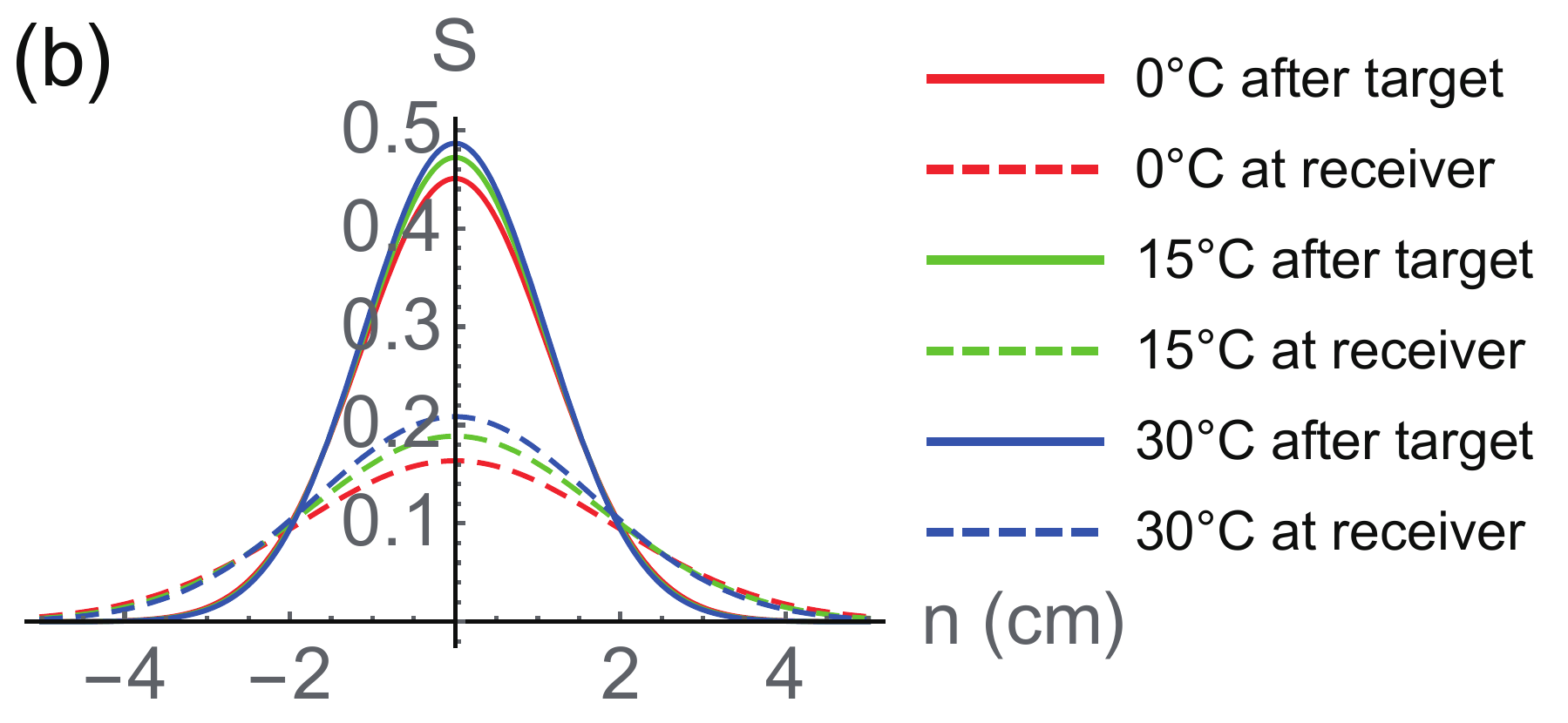}
		\label{fg:f6b}}
	\subfigure{
		\includegraphics[width=0.45\textwidth]{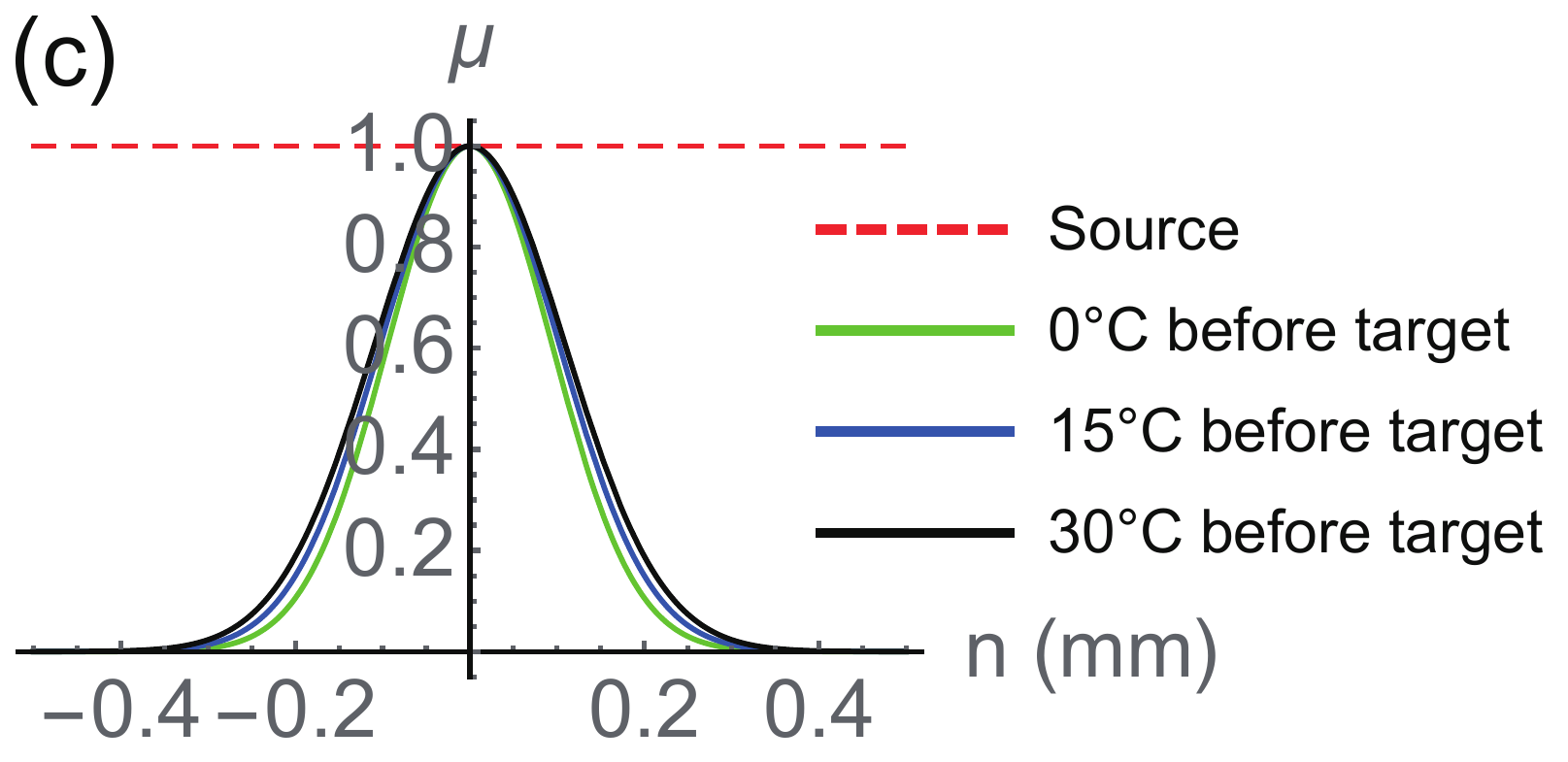}
		\label{fg:f6c}}
	\subfigure{
		\includegraphics[width=0.45\textwidth]{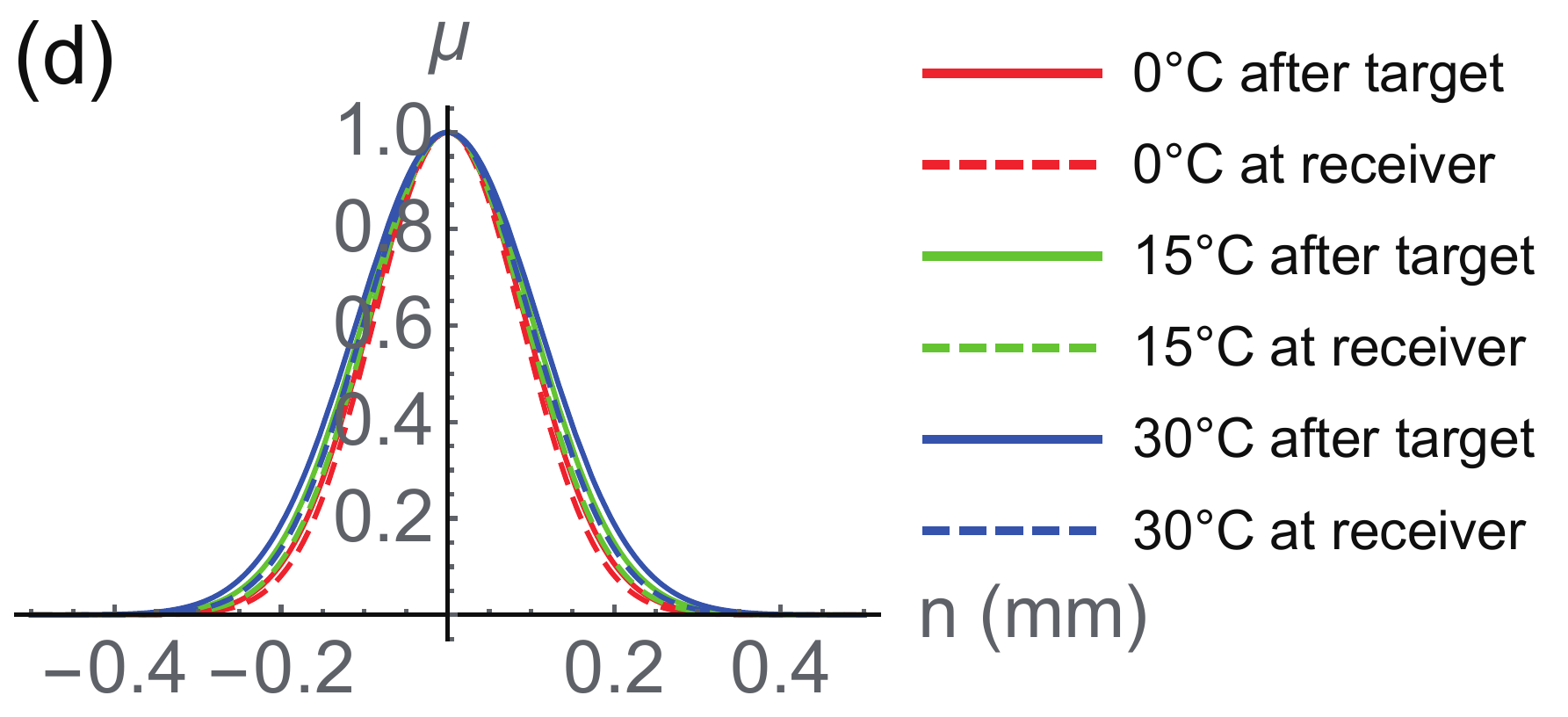}
		\label{fg:f6d}}
	\caption{The evolution of $S$ and $\mu$ in different $\left< T \right>$, where we set $\omega = -3$, $\varepsilon = 10^{-4} \rm m ^2 \rm s ^{-3}$, $\chi_{\rm{T}} = 10^{-5}\rm K ^2 \rm s ^{-1}$.}
	\label{fg:f6}
\end{figure*}

\begin{figure*}
	\centering
	\begin{minipage}[c]{.80\textwidth}
		\centering
		\subfigure[$\quad S$ varies with $\chi_{\rm{T}}$.]{
			\includegraphics[width=0.265\textwidth]{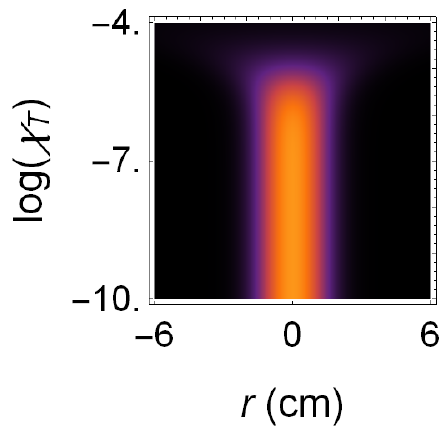}
			\label{fg:f7a}}
		$\ \ \ $
		\subfigure[$\quad S$ varies with $\varepsilon$.]{
			\includegraphics[width=0.265\textwidth]{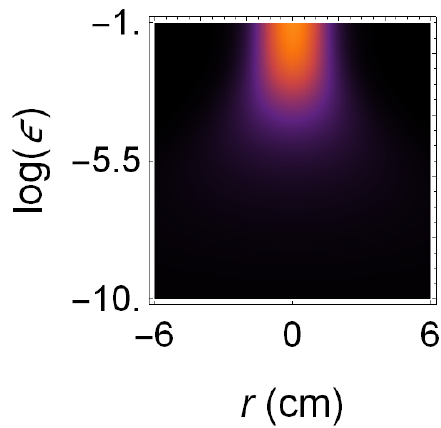}
			\label{fg:f7b}}
		$\ $
		\subfigure[$\quad S$ varies with $\omega$.]{
			\includegraphics[width=0.277\textwidth]{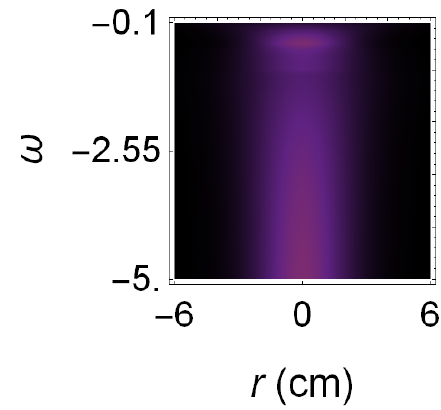}
			\label{fg:f7c}}\\
		\subfigure[$\quad \mu$ varies with $\chi_{\rm{T}}$.]{
			\includegraphics[width=0.265\textwidth]{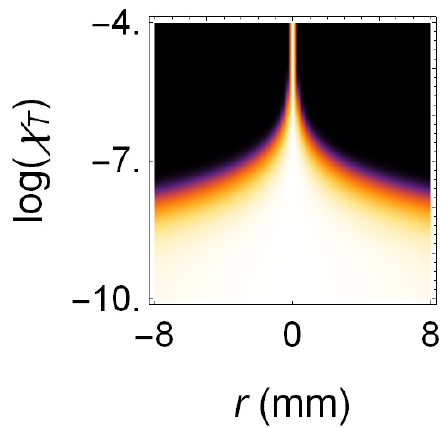}
			\label{fg:f7d}}
		$\ \ \ $
		\subfigure[$\quad \mu$ varies with $\varepsilon$.]{
			\includegraphics[width=0.265\textwidth]{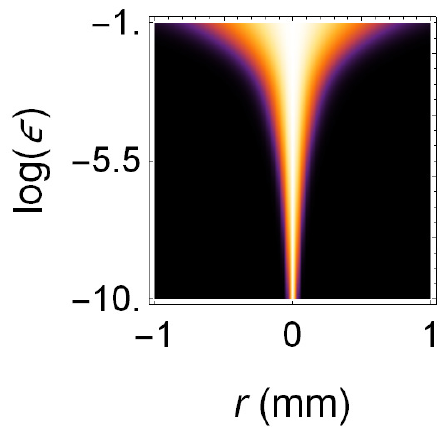}
			\label{fg:f7e}}
		$\ $
		\subfigure[$\quad \mu$ varies with $\omega$.]{
			\includegraphics[width=0.277\textwidth]{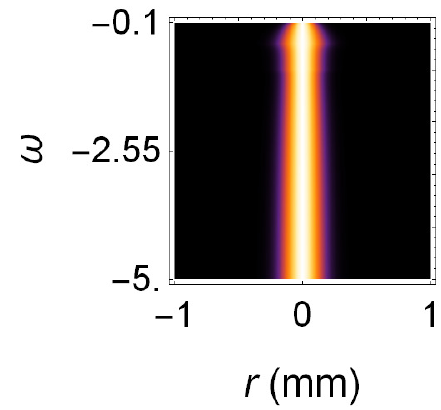}
			\label{fg:f7f}}
	\end{minipage}
	\begin{minipage}[c]{.1\textwidth}
		\subfigure{	
			\includegraphics[width=0.6\textwidth]{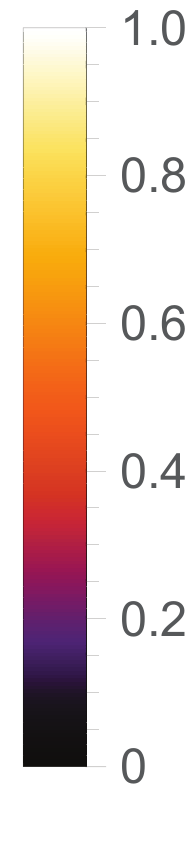}}
	\end{minipage} 
	\caption{
		$S$ and $\mu$ at receiver vary with $\chi_{\rm{T}}$, $\varepsilon$ and $\omega$. (a) and (d) vary $\chi_{\rm{T}}$ with $\omega = -3$, $\varepsilon = 10^{-4} \rm m ^2 \rm s ^{-3}$, $\left\langle T \right\rangle = 15^\circ \rm C$; (b) and (e) vary $\varepsilon$ with $\omega = -3$, $\chi_{\rm{T}} = 10^{-5}\rm K ^2 \rm s ^{-1}$, $\left\langle T \right\rangle = 15^\circ \rm C$; (c) and (f) vary $\omega$ with $\varepsilon = 10^{-4}\rm m ^2 \rm s ^{-3}$, $\chi_{\rm{T}} = 10^{-5}\rm K ^2 \rm s ^{-1}$, $\left\langle T \right\rangle = 15^\circ \rm C$.
	}
	\label{fg:f7}
\end{figure*}

In Fig. \ref{fg:f6}, we set $\omega = -3$, $\varepsilon = 10^{-4} \rm m ^2 \rm s ^{-3}$, $\chi_{\rm{T}} = 10^{-5} \rm K ^2 \rm s ^{-1}$, 
calculate $S(n,L)$ and $\mu (n,L)$ in different average temperatures $\left\langle T \right\rangle$. It shows that a lower $\left\langle T \right\rangle$ contributes to a stronger degeneration of $S$ and $\mu$. This degeneration narrows the curves of $\mu$, contributes to lower and broader curves of $S$. 
In Fig. \ref{fg:f7}, we calculate $S(r,L)$ and $\mu (r,L)$ at receiver with varied $\chi_{\rm{T}}$, $\varepsilon$ and $\omega$.
Figs. \ref{fg:f7a} and \ref{fg:f7d} are given with $\omega = -3$, $\varepsilon = 10^{-4} \rm m ^2 \rm s ^{-3}$, $\left\langle T \right\rangle = 15 ^\circ \rm C$, and $\chi_{\rm{T}}$ ranging from $10^{-10} \rm K ^2 \rm s ^{-1}$ to $10^{-4} \rm K ^2 \rm s ^{-1}$.
Figs. \ref{fg:f7b} and \ref{fg:f7e} are given with $\omega = -3$, $\chi_{\rm{T}} = 10^{-5} \rm K ^2 \rm s ^{-1}$, $\left\langle T \right\rangle = 15 ^\circ \rm C$, and $\varepsilon$ ranging from $10^{-10} \rm m ^2 \rm s ^{-3}$ to $10^{-1} \rm m ^2 \rm s ^{-3}$. 
In these four subfigures, the degeneration of $S$ and $\mu$ cased by turbulence is obvious. The degeneration is strengthened by the increased $\chi_{\rm{T}}$ or/and the decreased $\varepsilon$. 
In Figs. \ref{fg:f7c} and \ref{fg:f7f}, we set $\varepsilon = 10^{-4} \rm m ^2 \rm s ^{-3}$, $\chi_{\rm{T}} = 10^{-5} \rm K ^2 \rm s ^{-1}$, $\left\langle T \right\rangle = 15^\circ \rm C$, and vary $\omega$ from $-5$ to $-0.1$. The degeneration of $S$ varies weakly when $\omega < -2.5$, but when $\omega$ is around $-1.325$ and $\omega\to 0$, the variation of degeneration can be observed; the degeneration of $\mu$ increases slowly from $\omega = -5$ to $\omega = -1$, but decreases from $\omega = -1$ to $\omega = -0.5$, and finally increased quickly when $\omega \to 0 $. This complicated variation of degeneration comes from the complex variation of $\rho_{0}$ with $\omega$ (refers to Fig.\ref{fg:f4_b}).

Let us now summarize the results of the previous and this section, on the effect of various turbulence parameters on the beam evolution in the LIDAR systems:
\begin{itemize}
	\item Smaller values of average water temperature lead to a stronger turbulence, and, hence, stronger effect on the LIDAR's return;
	\item Larger values of $\chi_{\rm{T}}$ or smaller values of $\varepsilon$ lead to  degeneration to the LIDAR's performance, due to dependence  $\varepsilon^{-\frac{1}{3}}\chi_{i}$ in Eq. \eqref{eq:eq2};
	\item  Parameter $\omega$ has a somewhat complex influence on both, beam propagation and LIDAR performance, because of its nonlinear relations with other parameters (see Eqs. \eqref{eq10} and \eqref{eq07_3}).
\end{itemize}

\section{SUMMARY}

We have implemented the $4\times 4$ ABCD matrix method based on the extended Huygens-Fresnel integral for the analysis of optical beam propagation in the bi-static and mono-static (beyond the EBS area) LIDAR systems operating in the presence of underwater turbulence. By applying three transformations: for the beam passage from source to target, for its interaction with the target, and for its propagation from target to the receiver, sequentially, we illustrated how the second-order moments of the beam can be predicted and compared at the source plane, just before the target, just after the target and at the collecting lens of the receiving system. 

The general method is illustrated for the EM GSM beams, i.e., beams with arbitrary size, coherence and polarization states, having Gaussian profiles of average intensities and correlations. It can be extended to other sources with structured correlation functions as well. The $4\times 4$ ABCD matrix method also involves the knowledge of the coherence radius of the spherical wave that can be evaluated numerically from the corresponding structure function. For all calculations in this paper we have used a recently developed power spectrum \cite{ar01} that takes into account, in addition to other parameters, the average temperature in the water channel. We have found that when  other parameters of the source and of the channel are fixed, the coherence radius of the spherical wave slightly decrease with the increase of the average temperature. As a part of our calculations we have also elucudated that on a single-pass propagation in oceanic turbulence (in the absence of any optical elements or a target) the beam's spectral density and the degree of coherence are affected more for smaller average temperatures. This effect is in agreement with that for the scintillation index of a spherical wave that was previously calculated in Ref. \cite{ar01}.  

We have also carried out very detailed numerical analysis of the beam evolution in a bi-static LIDAR system, using, without loss of generality, a linearly polarized and fairly coherent source. Our results indicate that the underwater optical turbulence degenerates the spectral density and the degree of coherence as follows: a lower value of the average temperature, a larger value of the temperature dissipation rate, or a smaller value of kinetic energy dissipation lead to more beam broadening and lower coherence. We have also illustrated that other parameters of turbulence affect the beam in a very complex manner.





\section*{Appendix}
Following the process in Ref.\cite{ar01}, we show the values of viscosity, diffusivity and Prandtl/Schmidt number varying with environment in Table \ref{tab1}.
\begin{table*}
	\centering
	\caption{\bf Values of viscosity, diffusivity and Prandtl/Schmidt number in different values of temperature when the average salinity is $34.9\rm{ppt}$}
	\newcommand{\tabincell}[2]{\begin{tabular}{@{}#1@{}}#2\end{tabular}}
	\begin{tabular}{cccccccccc}
		\hline
		\tabincell{c}{Average salinity  $\left\langle S \right\rangle$  ($\rm{ppt}$)}  & \multicolumn{8}{c}{34.9} &\\
		\hline
		
		\tabincell{c}{Average temperature  $\left\langle T \right\rangle$  ($^{\circ}\rm C$)}  & 0 & 5 & 10 & 15 & 20 & 25 & 30 &\\
		\hline
		
		\tabincell{c}{Kinematic viscosity\\$\upsilon$ (${10^{ - 7}}{{\rm{m}}^2} \cdot {{\rm{s}}^{ - 1}}$)}  & 18.534 & 15.756 & 13.599 & 11.887 & 10.503 & 9.366 & 8.420 &\\
		\hline
		
		\tabincell{c}{Salinity diffusivity\\$\alpha _{\rm{S}}$ (${10^{ - 10}}{{\rm{m}}^2} \cdot {{\rm{s}}^{ - 1}}$)}  & 7.74 & 9.28 & 10.95 & 12.86 & 14.50 & 17.71 & 18.46 &\\
		\hline
		
		\tabincell{c}{Temperature Prandtl number\\ ${Pr _{\rm{T}}}$ (Dimensionless)}  & 13.349 & 11.182 & 9.516 & 8.205 & 7.155 & 6.301 & 5.596 &\\
		\hline
		
		\tabincell{c}{Salinity Schmidt number\\ ${Pr _{\rm{S}}}$ (Dimensionless)}  & 2393.2 & 1697.7 & 1241.6 & 924.3 & 724.3 & 528.8 & 456.1 &\\
		\hline
		
		\tabincell{c}{Kolmogorov microscale $\eta $ ($10^{-5}\rm{m}$)\\when $\varepsilon  = 1 \times {10^{ - 2}}{{\rm{m}}^2}{{\rm{s}}^{ - 3}}$} & 15.885 & 14.063 & 12.593 & 11.384 & 10.375 & 9.521 & 8.790 &\\
		\hline
	\end{tabular}
	\label{tab1}
\end{table*}

\section*{References}

\end{document}